\documentstyle[12pt]{article}
\newcommand{\JJ}{$J_1-J_2$ model}

\newcommand{\be}{\begin{equation}}
\newcommand{\ee}{\end{equation}}
\begin{document}
\begin{titlepage}
\begin{center}
{\Large \baselineskip6pt {\bf $\bf J_1-J_2$ quantum Heisenberg
antiferromagnet: Improved
spin-wave theories versus exact-diagonalization data} \\}
\vspace{1.5cm}
\normalsize
{\it \baselineskip6pt N.B. Ivanov$^*$ and J. Richter \\
\vspace{0.5cm}
\baselineskip6pt Institut f\"ur Theoretische Physik \\
Otto-von-Guericke Universit\"at Magdeburg\\
Postfach 4120\\
D-39016 Magdeburg\\
Germany\\}
\end{center}

\vspace{2cm}

short title: {\bf $\bf J_1-J_2$ quantum Heisenberg
antiferromagnet}

\vspace{1cm}


\vspace{0.9cm}

\baselineskip6pt {$^*$ Permanent address: Georgi Nadjakov  Institute  for Solid
State
Physics, Bulgarian Academy of Sciences, 72 Tzarigradsko chaussee blvd.,
1784 Sofia, Bulgaria}
\end{titlepage}
\newpage

\vspace{1.5cm}

{\bf Abstract}\\

We reconsider the results cocerning the  extreme-quantum
$S=1/2$ square-lattice Heisenberg antiferromagnet with frustrating
diagonal  couplings  ($J_1-J_2$ model) drawn from  a
 comparison with exact-diagonalization data. A combined approach
using also some intrinsic features of the self-consistent
spin-wave theory leads to the conclusion that the theory strongly
overestimates the stabilizing role of quantum flutcuations in
respect to the N\'{e}el phase in the extreme-quantum case $S=1/2$.
 On the other hand, the analysis
implies that the N\'{e}el phase remains stable at least up to the
limit $J_{2}/J_{1} = 0.49$
 which is pretty larger than some previous estimates.
In addition, it is argued that the spin-wave ansatz predicts
the existence of a finite range ( $J_{2}/J_{1}<0.323$
in the linear spin-wave theory ) where
the Marshall-Peierls sigh rule survives the frustrations.
\newpage

\section{Introduction}\
   The square-lattice Heisenberg antiferromagnet with  antiferromagnetic
   next-nearest-neighbor couplings (\JJ) produces a
simple and, at the same time, an important example of a frustrated
quantum spin system. The model is defined by the Hamiltonian
\be\label{h}
H = J_1\sum_{\langle i,j \rangle} {{\bf S}_i{\bf S}_j} +
J_2\sum_{[i,j]} {{\bf S}_{i}{\bf S}_j} \hspace{0.5cm}, \hspace{2cm} J_1,J_2>0,
\ee
where the symbols $<i,j>$ and $[i,j]$ mean  that  the  summations run over  the
nearest-neighbor  and  next-nearest-neighbor  (diagonal)   bonds,
respectively. In what follows we put $J_1\equiv 1$,  $ \alpha \equiv
J_2/J_1$.

Presently little is  known
about the ground-state properties of this model. In the classical
limit $S=\infty$ the \JJ\ has two phases: if $\alpha < 1/2$ ,the ground  state
is a two-sublattice N\'{e}el state, whereas, if $\alpha > 1/2$, the
 four-sublattice antiferromagnetic state is stable. At the classical transition
point $\alpha=1/2$ the model is characterized by a  great  degree
of classical degeneracy: all states with zero  elementary-plaquette
spins are energetically preferable. The quantum fluctuations,  however,
can drastically change this picture. In general,  they  are
determined by the microscopic structure of the  model,  and
are expected to increase as $S$ approaches  the  extreme-quantum
limit $S=1/2$, and/or the frustration becomes stronger. Already a simple
linear spin-wave analysis reveals  such  a  tendency [1].
In addition, the latter theory predicts the existence of  a  finite  range
around the classical phase boundary $\alpha=1/2$ where the classical long-range
magnetic  order  is  completely
destroyed (for arbitrary $S$). However, the next-order  terms  in the
 large-$S$  expansion
show logarithmic divergencies [2] connected to an additional softening
of the spectrum at $\alpha=1/2$,
 thus making the first-order  predictions, at least, questionable. This
 situation is characteristic for most of the studied frustrated
 models. An open question is how to reconstruct the standard spin-wave
 expansion in order to avoid the mentioned difficulties.
The Hartree-Fock  type
theories [2-4], which could in principle serve as a starting point
for  a systematic expansion, predict  a  first-order  phase
transition   between
the magnetically ordered phases without any intermediate phase.  This
picture is connected  with  the  predicted  stabilizing  role  of
quantum fluctuations in respect to the two-sublattice N\'{e}el
order. Presently, however, it is not clear  if  these  conclusions
are characteristic, at least  qualitatively,  for  the  extreme-quantum
system $S = 1/2$ as well.

Concerning the $S=1/2$ case, at least  two  important  issues,
related to the ground-state phase  diagram,  remain  unsettled: (i) the
nature of the magnetically  disordered  phase, if any,  in  the  strongly
frustrated  region; (ii) the  location  of  the  phase-transition
boundary. The magnetically disordered spin-Peierls dimer state is
preferable in a number of  studies: 1) series  expansions  around
dimer states [5], 2) $1/N$-expansion technique [6], 3) bond-operator
techniques [7], 4) effective-action approaches leading to quantum
nonlinear $\sigma$-models [8], 5) numerical exact-diagonalization
data [9,10].
However, each of the mentioned methods has its own defects,  so
that some other states (e.g., the chiral  states [10,11,12]) seem to
be possible candidates, as well.

With regard to  the  location  of  phase  boundary,  here  the
estimates run in the large interval from $\alpha_{c} \approx 0.15$ to
$\alpha_{c} \approx  0.6$. The bound $\alpha_{c} \approx
 0.15$ was obtained [13] by use  of $\sigma$-model considerations combined
with Schwinger-boson mean-field results for $S=1/2$. On  the  other
hand, the largest estimate $\alpha_{c} \approx 0.6$ is  characteristic
 for  the
self-consistent theories [2-4]. Series of  studies give values which
are near the point $\alpha_{c}=0.4$ [1,9,10].

The outlined ambiguity signals of a lack  of  reliable  descriptions
even  in  the  weakly  frustrated  region  where  the  two-sublattice
 N\'{e}el phase is expected to be stable. Concerning the spin-wave
  theories, a way to test their quality gives  the  comparison
with numerical exact-diagonalization data. For the  \JJ\,
the first steps in this direction were made by Hirsch and Tang [14]
based on Takahashi's idea [15] for a constrained spin-wave theory in
low dimensions. These authors indicated that their theory systematically
overestimates the effect of frustration in  destroying
the  antiferromagnetic  correlations  ($N= 10,16,26$). Recently,
Ceccatto, Gazza, and Trumper [16] have continued this line by use of
Takahashi's self-consistent approach [17]   adapted to the  frustrated
model [4]. A remarkable agreement with the exact results for a  number
of lattices ($N= 10,20,26$) was indicated, excluding, however,
the most symmetrical lattice $4 \times 4$.

In this paper we study the extreme-quantum system $S=1/2$
and show that the existing exact-diagonalization results
in combination with some intrinsic properties of the self-consistent
theory lead to the following conclusions: (i) The classical
N\'{e}el state is stable at least up to the limit $\alpha^{*} =0.49$.
 Notice that the estimate is quite
larger than the result
$\alpha \approx 0.4$ mentioned above; (ii) The self-consistent
spin-wave theory overestimates the role of quantum fluctuations
in stabilizing the N\'{e}el state.  This last conclusion
also differ from previous considerations relying on a
comparison with exact-diagonalization results for less
symmetrical lattices $N= 10,20,26$ when the theory
indeed gives exellent results.

\section{Comparison of the theory
with exact-diagonalization data}\

\subsection{ Fitting to the exact-diagonalization
results: $N=16$ lattice}

In reconsidering the previous exact-diagonalization data,  it is easy to
see a well-pronounced tendency, i.e., the self-consistent spin-wave
theory gives  good correlators for a number of
less-symmetrical
lattices $(N=10,20,26)$, whereas  the  most  symmetrical
4x4 lattice is aside from this tendency.  On  the  other
hand, Hirsch-Tang's theory  overestimates  the
effect of frustration for all lattices ( including  the 4x4
  lattice).  In
order  to further check this  observations,  we  present  here  new
exact-diagonalization results for $N=18$ and $N=24$ $(6 \times 4)$
lattices, Figs.1,2. The
lattice $N=18$, belonging to the class of lattices $N=10,20,26$,
 is expected to suppress more symmetrical fluctuations,
including the four-sublattice state fluctuations (because  $N/2=9$
is odd), whereas the $6 \times 4$ lattice is rather closer to the $4
\times 4$ lattice. It is seen that  the tendency is conserved: For the
 $N=18$ site lattice the self-consistent theory
 practically reproduces the exact data
up to $\alpha \approx 0.4$, whereas for the $N=24$ lattice
it evidently underestimates the role of frustrations starting
from $\alpha =0$. On the other hand, Hirsch-Tang's
consideration [14], which
does not take into account quantum fluctuations,
predicts less sublattice magnetizations (Fig.1) in both kind of
lattices. Notice that the latter theory leads to
 completely wrong correlators (they are not presented in Fig.2)
 for the discussed lattices. Therefore, we suggest that the
symmetrical lattices $N=16,24$ reproduce in a more adequate way the
main properties of the thermodynamic limit.

In what follows we address  the most symmetrical $4 \times 4$ lattice.
   The scaling parameter $U=f/g$ (see Ref.4)  appears  in  the
self-consistent spin-wave theory
through a Hartree-Fock decoupling of  the  quartic  terms  in  the
Hamiltonian. Within  the   theory,  $U$  is  given  by  the
self-consistent equations. In principle, one can use $U$ as a
variational parameter in the spin-wave ansatz
\be\label{psisss}
\psi_{S}  \hspace{4pt} \sim \hspace{4pt}{\it P}
\exp {\Big(\! \sum_{\bf k}{\!
' w_{{\bf k}} \hspace{4pt}
\hat{a}^+_{{\bf k}} \hat{b}^+_{-{\bf k}}}  \Big)} \hspace{7pt}
|N\acute{e}el\rangle \hspace{0.5cm} .
\ee
Here $|N\acute{e}el\rangle$ is the classical N\'{e}el state. The  weight
 factors  $w_{\bf k}$ are defined by $w_{\bf k} = v_{\bf k} / u_{\bf k}$
    ,  $v_{\bf k}$ and  $u_{\bf k}$ being the  well-known
Bogoliubov coefficients; {\it P} is a projection operator, and
the  prime means that the sum runs over the small Brillouin zone. This
variational state is studied in Ref.[19]. Here we treat $U$ as a fitting
parameter obtained from a requirement for best fitting between the
sublattice magnetization $M_{s}^2$, as obtained from the theory, and the
exact-diagonalization results. The reasoning for such a consideration
comes from the following observations:

(a) From Takahashi's condition $S_z=0$, when applied in the limit $N= \infty$,
 one can directly deduce the following scaling
relation connecting the sublattice magnetizations in the linear
spin-wave approximation $m_{0}(\alpha)$ (when $U=1$) and in the
self-consistent theory $m(\alpha)$:
\be\label{m}
 m(\alpha)= m_{0}( \alpha U), \hspace{0.5cm} N= \infty .
\ee
It is interesting to notice that the above scaling relation does not
explicitly depend on the site spin $S$ (apart from a
trivial  linear  term ). The implicit dependence is
hidden in the scaling factor $U$  which for $\alpha =0$ reads:
\be\label{u1}
U = \frac{1-0.102/2S+O[{(2S)}^{-2}]}{1+0.158/2S+O[{(2S)}^{-2}]},
\hspace{0.5cm} N= \infty.
\ee
For $S=1/2$ one gets $U=0.775$. The next-order term slightly
diminishes the latter number. The self-consistent theory
predicts a monotonic desrease of $U$ versus $\alpha$ in
the whole range where the classical Neel state is stable ($U \approx
0.6$ at the phase transition point $\alpha_c \approx 0.62$). If one takes
the self-consistent theory as a starting point for a systematic
perturbation expansion, one can hardly expect any drastic qualitative
change in
the behavior of $U$ versus $\alpha$. The arguments are as follows:
First, in this approximation the unphysical modes due to the
degeneracy of the classical ground state at $\alpha =1/2$
acquire gaps, so that some of the problems concerning the standard spin
wave approach here are resolved. Second, the denominator in Eq.(4) is
just the rescaling factor of the spin-wave velocity, which is
expected (also from other methods) to be slightly $\alpha$-dependent
and finite at the phase boundary.

(b) A direct calculation of the scaling factor $U$ for $N=\infty$
and $N=16$
gives practically the same function $U(\alpha)$.
 In other words, the $N=16$ lattices is large
enough in respect to this quantity, as it should be expected,
because the factor $U$ is a ratio of two short-ranged bosonic
correlators. This observation will be used to get
information concerning the $N=\infty$ system.
The results coming from an exact fitting of the theoretical and
exact-diagonalization functions $M_{s}^2$  are as follows:

(i) The resulting scaling factor $U$ is approximatetly
$\alpha$-independent  with a value close to
the one predicted by Eq.(4).
(ii) It is seen a remarkable fit to the exact  correlators (Figs.3)
practically in the whole interval up to $\alpha \approx 0.45$.
 The misfit for $\alpha > 0.45$ is easily indicated in the  ground-state
energy because  the small overestimates  for  the  short-range
correlators, noticed  in Fig.3a, are summed.  Nevertheless,
we have checked that the energy is approximately unchanged by the
fitting up to $\alpha \approx 0.45$.

 Based on the argument (b) and the suggestion that the $N=16$ lattice
better reflects the $N= \infty$ limit (as compared to less
 symmetrical lattices), one can predict the same picture in the
 thermodynamic limit: Namely, the monotonic decrease in the
 scaling factor $U$ vs $\alpha$ should become smoother (as it is
for $N=16$) in a
 more refined approximation starting from the discussed self-consistent
 theory. In particular, the phase boundary $\alpha_c
 \approx 0.62$ should drastically move towards smaller $\alpha$ (
because $U$ vs $\alpha$ is approximately constant according to
the $N=16$ fitting). The last two
equations, combined with the hypothesis for a smooth
 decrease of the
 scaling factor $U$ vs $\alpha$, give the following lowest
limit where the N\'{e}el phase becomes unstable:
$ \alpha^{*} > 0.49$, $S=1/2$. As a matter of fact, this estimate can be
slightly increased if one takes  the next-order approximation in Eq.(2).

To summarize, a combined approach relying on a comparison with
exact-diagonalization data and some intrinsic features of
the self-consistent spin-wave theory ( the scaling relation (3),
the smooth monotonic decrease of $U(\alpha)$, and the short-range
character of the scaling factor $U$ ) lead to the conclusion that
the theory,when applied to the extreme-quantum system $S=1/2$,
overestimates the stabilizing role of quantum fluctuations. In
addition, the same analysis predicts a lowest limit $\alpha^{*} = 0.49$
where the N\'{e}el state
is destroyed which is quite larger than the
previous estimate $\alpha \approx 0.4$
based on the linear spin-wave theory [1,14] and on the
$N=36$ exact-diagonalization results [10].

An additional understanding of the features of the
self-consistent approximation can be obtained from the spin ansatz
(2). Here we address
the square of the sublattice  magnetization  $M_s^2$,
Fig.4. The exact $M_s^2$ is  compared  to: (i) self-consistent theory
; (ii) $U=1$, i.e., linear theory; (iii) the ansatz
$\psi_{\frac{1}{2}}$, Eq.(2), with $U=1$.
 It is obvious that the function $M_s^2$ (calculated
with the ansatz $\psi_{1/2}$, $U=1$) strictly follows
 the form of
the exact function $M_s^2(\alpha)$ in a large region up to $\alpha=0.5$.
The main  difference
between the theory and $\psi_S$ for $U=1$ lies in the fact that
the variational function  does  not
contain unphysical states. Therefore, the increasing misfit in $M_s^2$
(and in the other correlators in  Hirsch-Tang's theory)
is predominantly connected with the enhanced role of the
unphysical bosonic states in the frustrated system (notice that already
the
linear spin-wave approximation in the pure $\alpha=0$ system gives
a good estimate for the reduced site spin $m_0=0.303$).
\vspace{0.5cm}

\subsection{Violation of the Marshall-Peierls sign rule}

Recent exact numerical  diagonalization studies
   of small lattices [19] show that  the  ground-state  wave
   function of the $S=1/2$ \JJ\ violates the Marshall-Peierls
sign rule for sufficiently large $\alpha$. Here we  present results
which are based on the spin-wave ansatz (2).
 Originally,  the  mentioned  rule
had been  proved  for  bipartite
lattices with nearest-neighbor interactions [20]. The latter says  that
the ground-state wave function  of  $S=1/2$  Heisenberg  antiferromagnet
 reads
\be\label{srule}
 \psi =\sum_{n} {(-1)}^{p_n} a_{n} |n \rangle, \hspace{1cm} a_{n}>0,
 \ee
where $|n \rangle$ is an Ising state, $p_{n}$ being the number  of,
say,  up-spins living on, say, A-sublattice. Notice that the  proof does
not  work  for  a  system  with  antiferromagnetic  next-nearest-neighbor
 (diagonal) couplings. As a matter of fact,  this rule is
violated, as mentioned above, in the $4 \times 4$ lattice
provided   the frustration is strong enough.

Firstly, let us rewrite the spin-wave ansatz (2)  in the
form:
\be\label{psi12}
\psi_{\frac{1}{2}} \hspace{4pt} \sim \hspace{4pt} \prod_{{\bf R} , {\bf r}
 \atop {{\bf R} \in A}}{\Big[ 1-w({\bf r})S_{\bf R}^{+} S_{\bf R+r}^{-}
\Big]}
|N\acute{e}el\rangle \hspace{0.5cm},
\ee
where the pairing function $w({\bf r})$ is defined by
\be\label{w}
w({\bf r}) \, = \, \frac{2}{N} \sum_{\bf k}{\! ' w_{\bf k} \cos{\bf kr}}
\hspace{0.5cm}.
\ee
The vector ${\bf r}$ in Eqs.(6,7) connects  sites from  different
sublattices.

 From the structure of the ansatz
 it  is
clear that the sign rule breaks if,  and  only  if,  the  pairing
function  $w({\bf r})$ changes its sign for some vector ${\bf r}$
 connecting  two spins which live on different sublattices.
 For the  $4 \times 4$  lattice this is just the vector
 ${\bf r} = \hat{\bf x}+2 \hat{\bf y}$ (and  the  related  by  symmetry
vectors on the lattice).  The pairing  function
$w( \hat{\bf x}+2 \hat{\bf y})$ vs $\alpha$ is presented in Fig.5.
For $U=1$, $w( \hat{\bf x}+2 \hat{\bf y})$ changes  sign
at a point practically coinciding with the related $N = \infty$ limit,
$\alpha_M = 0.323$ (this is another indication that
this symmetrical lattice covers quite well some characteristic features
of the infinite system). This
characteristic point preceeds the
instability point $\alpha^{*}$ . These  observations  were
based on the spin-wave ansatz (2). The predicted weight of Marshall states
(Ising
states fulfilling the rule) vs $\alpha$  is in agreement with the
exact result presented in Fig.6. Quite surprisingly, a recent study of the
same problem
for ground states with larger total-spin quantum numbers,
$S_{total}=1,2$, and
$3$, [21] indicates a sharp increase of the weight of non-Marshall states
near the point $\alpha \approx 0.52$ which is pretty close to limit
$0.49$ found in the present consideration ( see also Ref. 22).

\section{Concluding remarks}
  The analysis presented above was based on a combined approach
  using exact-diagonalization data for small
  lattices and some intrinsic features of the self-consistent
spin-wave theory. It was directed
towards checking the predictions of the latter
theory for the extreme-quantum system $S=1/2$. It was
 found a stable tendency, namely, the theory excellently fits
 to the exact data for less symmetrical clusters ($N=10, 18, 20, 26$),
 whereas for $N=16$ (and also $N=24$) this approach evidently
 underestimates the effect of
 the frustrations.  At the same time, Hirsch-Tang's theory, which
 does not take into account spin-wave interactions, systematically
 overestimates the role of the frustrations for each of
 the mentioned lattices. This tendency probably means that some
 more symmetrical (e.g., four-spin) correlations,
 which are  suppressed  in  less-symmetrical
lattices, are not properly taken into account in the self-consistent
approach.
That is way the main conclusions are drawn from the
 comparison with the
 $N=16$ lattice which is suggested to reproduce better the properties
 of the $N= \infty$  model. The comparison gives an estimate $\alpha^{*}
 > 0.49$ for the location of the instability point $\alpha^{*}$,
which is  higher as compared to some previous results (Refs. 1, 8, 10). The
 analysis also shows that in a more refined approximation, using as a
 starting point the discussed theory, $\alpha^{*}$ should move
 towards smaller values of $\alpha$  (as a matter
 of fact, the fitting based on the $N=16$ lattice predicts an
 approximately constant rescaling factor $U$ up to the phase boundary,
  which would mean, if applied to the $N=\infty$ system, that the
  instability of the N\'{e}el phase should be close to the estimate
  $0.49$. Clearly, one needs some additional arguments in favor
  of such a suggestion ( see, e.g., Ref. 22).
  Finally, it was shown that the spin-wave projected ansatz
    predicts  a finite region ( $\alpha < 0.323$ in linear spin-wave
    theory) where the Marshall-Peierls sign rule is fulfilled in the
    frustrated $N=\infty$ model. For $N=16$, this result was shown
     to be in accord with the
   exact-diagonalization data.           \\
\vspace{0.2cm}

{ \Large {\bf Acknowledgements}} \\

   N.I.  is  grateful  to  Institut  f\"{u}r  Theoretische  Physik,
   Universit\"{a}t  Magdeburg,  for  hospitality  during
    his  stay in Magdeburg where a part of the work was accomplished.
     This research
was  supported  by  Deutsche  Forschungsgemeinschaft,  Grants
Ri 615/1-1, 436 BUL 112-13-93, and  Bulgarian Science Foundation,
Grant  $\Phi$2/91.
\newpage
{\Large {\bf References}} \\[0pt]
[1] \hspace{0.1cm} Chandra P and  Doucot B, Phys.Rev.{\bf B38},9335(1988);

\hspace{2pt} Oguchi T and Kitatani H, J.Phys.Soc.Jpn.{\bf 59},
3322(1990).\\[0pt]
[2] \hspace{0.1cm} Chubukov A, Phys.Rev.{\bf B44},392(1991);

 \hspace{2pt} Bruder C and Mila F, Europhys.Lett.{\bf 17}, 463(1992).\\[0pt]
[3] \hspace{0.1cm} Nishimori H and Saika Y, J.Phys.Soc.Jpn.{\bf 59},
4454(1990);

 \hspace{2pt}  Barabanov A F  and  Stary'h  O A,
  Pis'ma  Zh.Eksp.Teor. Fiz.{\bf 51}, 271
  (1990)

\hspace{2pt}  [JETP   Lett.{\bf 51}, 311(1990)];

 \hspace{2pt} Mila F, Poilblanc D, and
  Bruder C, Phys.Rev.{\bf B43}, 7891(1991);

 \hspace{2pt} Bergomi L and  Jolicoeur  Th,
   J. Phys. (France) {\bf 2}, 371(1992).\\[0pt]
[4] \hspace{0.1cm} Xu J H  and  Ting C S , Phys.Rev.{\bf B42}, 6861(1990);

\hspace{2pt} Ivanov N B  and Ivanov P Ch , ibid. {\bf B46}, 8206(1992).\\[0pt]
[5] \hspace{0.1cm} Gelfand M.P., Singh R.R.P.,and Huse D.A., Phys.Rev.{\bf
  B40}, 10801(1989).\\[0pt]
[6]  \hspace{0.1cm} Read N and  Sachdev S, Phys.Rev.Lett.{\bf
62},1694(1989);

\hspace{2pt} Nucl.Phys.  {\bf B316}, 609(1989).\\[0pt]
[7] \hspace{0.1cm} Sachdev S and Bhatt R N, Phys.Rev.{\bf B41},
9323(1990);

\hspace{2pt} Chubukov A V and
  Jolicoeur Th, ibid.{\bf B44}, 12050(1991).\\[0pt]
[8] \hspace{0.1cm} Einarsson T and Johannesson H, Phys.Rev.{\bf
B43},5867(1991).\\[0pt]
[9] \hspace{0.1cm} Dagotto E and  Moreo A, Phys.Rev.Lett.{\bf 63},
2148(1989);

\hspace{2pt} Singh R R and Narayanan R, ibid.{\bf 65}, 1072(1990).\\[0pt]
[10] Schulz J and Ziman T A L , Europhys.Lett.{\bf 18}, 355(1992).\\[0pt]
[11] Wen X G ,  Wilczek F,  and   Zee A,   Phys.Rev.{\bf
   B39}, 11413(1989).\\[0pt]
[12] Richter J, Gros C, and Weber W, Phys.Rev.{\bf B44}, 906(1991);

\hspace{2pt}   Richter J, ibid.{\bf B47}, 5794 (1993).\\[0pt]
[13] Einarsson T, Fr\"ojdh P,  and  Johannesson H,
Phys.Rev.{\bf B45}, 13121
   (1992).\\[0pt]
[14] Hirsch J E  and Tang S , Phys.Rev.{\bf B39}, 2887(1989).\\[0pt]
[15] Takahashi M, Phys.Rev.Lett.{\bf 58}, 168(1987).\\[0pt]
[16] Ceccatto H A , Gazza C J ,  and   Trumper A E ,   Phys.Rev.{\bf B45}, 7832
   (1992).\\[0pt]
[17] Takahashi M, Phys.Rev.{\bf B40}, 2494(1989).\\[0pt]
[18] Mermin N D and Wagner H , Phys.Rev.Lett.{\bf 22}, 1133(1966).\\[0pt]
[19] Retzlaff K, Richter J, and Ivanov N B, Z. Phys.{\bf B93},
21(1993).\\[0pt]
[20] Marshall W, Proc.Roy.Soc.{\bf A232}, 48(1955).\\[0pt]
[21] Richter J, Ivanov N B, and Retzlaff K, Europhys.Lett.(1994), in
press.\\[0pt]
[22] {\small After the present paper has been submitted for publication
we received a preprint (Gochev I, Phys.Rev.{\bf B}, (1994), to be
published) where the same problem is analyzed by an
 expansion for which Takahashi's
approximation appears as a zeroth  order. Our conclusions, drawn from the
exact-diagonalization results and  more
intuitive considerations, seem to be in an exellent agreement with the
latter author who has found $\alpha^{*} \approx 0.52$ (
the collinear state is found to loose stability at $\alpha \approx
0.57$).}

\newpage
{\Large {\bf Captions of figures}} \\
{\bf Fig.1}: The square of sublattice magnetization vs $\alpha$ for
$N=18$  and $N=24$ lattices. $U=1$ ($U_{sc}$) curve corresponds to
Hirsch-Tang's (self-consistent) theory. The points  are  the
exact-diagonalization data.\\
{\bf Fig.2}: Spin-spin correlators $<{\bf S_0}{\bf S_R}>$ (${\bf R} =
n \hat{\bf x}+m \hat{\bf y}$) for the $N=18$
lattice in the self-consistent theory. The respective $U=1$
 curves, which are not drawn here, are very bad. The points are the
 exact-diagonalization data.\\
{\bf Fig.3}: The  correlators $<{\bf S_0}{\bf S_R}>$ (${\bf R} =
n \hat{\bf x}+m \hat{\bf y}$) for the $N=16$  lattice. The
solid lines are the results of the fitting. The other notations
are the same as
in Fig.1 and Fig.2.\\
{\bf Fig.4}: The square of sublattice magnetization  vs
$\alpha$  for
the $N=16$ lattice. The curves
1 and 2 represent the spin-wave results for  $U=1$ and
$U_{sc}$, respectively. The curve 3 is
calculated with the spin-wave ansatz, Eq.(2), for $S=1/2$. The points are the
exact-diagonalization results.\\
{\bf Fig.5}: The pairing function $w( \hat{\bf x}+2 \hat{\bf y} )$,
as defined by
Eq.(7), vs $\alpha$, $U=1$.
$w(\hat{\bf x}+2 \hat{\bf y})$ vanishes at $\alpha_M = 0.323$
for $U=1$   in the
thermodynamic limit $N = \infty$.\\
{\bf Fig.6}:  The  weight  of  Marshall  states  vs  $\alpha$.
The curves 1 and 2 correspond to $U=1$ and $U_{sc}$, and
are calculated with $\psi_{\frac{1}{2}}$, Eq.(7).
  The  curve 3 represents the exact-diagonalization results, $N=16$.\\
\end{document}